\documentclass[journal]{IEEEtran}

\ifCLASSINFOpdf
   \usepackage[pdftex]{graphicx}
\else
   \usepackage[dvips]{graphicx}
\fi
\usepackage{autobreak}
\usepackage{booktabs}
\usepackage{multirow}
\usepackage{multicol}

\usepackage{amsmath}
\usepackage{algorithm}

\usepackage{algorithmic}

\usepackage{array}

\begin{document}
%
\title{Quantifying Independence Redundancy in Systems: Measurement, Factors, and Impact Analysis}
%
%
%

\author{Hong~Su
\IEEEcompsocitemizethanks{\IEEEcompsocthanksitem H. Su is with School of Computer Science, Chengdu University of Information Technology, Chengdu, China, 610041.\protect\\
E-mail: suguest@126.com
}
\thanks{}}

\markboth{Journal of \LaTeX\ Class Files,~Vol.~14, No.~8, August~2015}%
{Shell \MakeLowercase{\textit{et al.}}: Bare Demo of IEEEtran.cls for IEEE Journals}

\maketitle

\begin{abstract}
  Redundancy represents a strategy for achieving high availability. However, various factors, known as singleness factors, necessitate corresponding redundancy measures. The absence of a systematic approach for identifying these singleness factors and the lack of a quantifiable method to assess system redundancy degrees are notable challenges.
  In this paper, we initially present methodologies to evaluate system redundancy, specifically quantifying independent redundancy in complex systems. This approach considers the interactions among various factors that influence redundancy, treating different factors as distinct dimensions to comprehensively account for all potential impact factors.
  Additionally, we propose methodologies to calculate the Independent Redundancy Degree (IRD) when combining or removing system components, offering insights into system resilience during integration or separation.
  Furthermore, we broaden the scope of known singleness factors by exploring time and space dimensions, aiming to identify additional related singleness factors. This process helps us pinpoint critical system aspects that necessitate redundancy for enhanced fault-tolerance and reliability.
  The verification results underscore the influence of different dimensions and reveal the significance of addressing weak dimensions for enhancing system reliability.
\end{abstract}


\IEEEpeerreviewmaketitle

\section{Introduction} \label{section_introduction}
The evolution of computing technologies has led to an increasing demand for highly reliable and resilient systems that can provide uninterrupted services \cite{ma2019high} \cite{elbamby2019wireless}. In the early days of computing, a single hardware component and software program were sufficient to execute tasks. However, the susceptibility of both hardware and software to errors and failures posed significant challenges to system reliability. A single hardware or software outage could render the entire system incapable of providing services. To address this issue, redundancy techniques were introduced to enhance system robustness.

\begin{figure}
  \centering
  \includegraphics[width=3.5in]{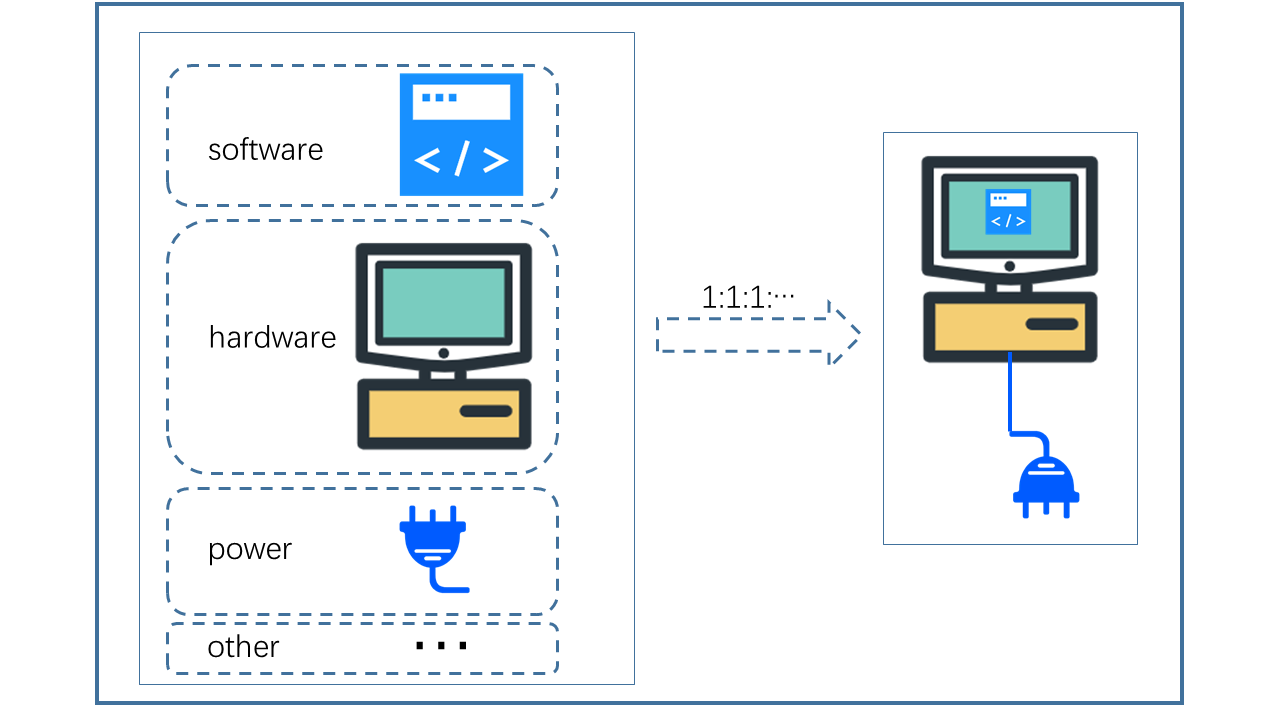}
  \caption{Components of a non-redundant system. In this configuration, software, hardware, and other elements (on the left) are interconnected on a one-to-one basis to constitute a system (on the right).}
  \label{singleComputerSingleSoftware}
\end{figure}

Redundancy strategies have become instrumental in improving system reliability. One common approach is the use of P2P (peer-to-peer) hardware or network, where multiple redundant hardware components are employed to mitigate random errors. While this approach offers a level of fault tolerance, it is not foolproof. For example, if all redundant hardware components are placed in the same physical location and a catastrophic event such as an earthquake occurs, the entire system may still fail. To address this limitation, the concept of geographical redundancy emerged, where redundant hardware is distributed across different cities or even countries to ensure system resilience \cite{li2019cost}.

Similarly, redundant software solutions have been proposed to address issues related to single software failures. In cases where a software contains a critical bug that causes it to crash when processing specific data, deploying multiple copies of the software on separate hardware does not solve the problem, as all copies will encounter the same error. To overcome this, independent redundant software instances are used to process user services. If one software encounters a fatal bug, the other independent instances can continue to function normally.

While redundancy techniques have made significant progress, several crucial questions remain unanswered. These inquiries encompass comprehending the interconnections between various redundancy methods, assessing the attainable extent of system redundancy. It's imperative to ascertain the feasibility of establishing a completely redundant system that can deliver services seamlessly under various conditions, encompassing hardware, software, and communication network failures. Moreover, there is a need to develop robust metrics for quantifying the degree of redundancy within a system and to formulate methodologies to effectively compare redundancy levels across different systems.

In this paper, we aim to address these questions and provide a formal definition of redundancy and independent redundancy. We will analyze the relationships among various redundant methods and their corresponding impact factors. Additionally, we will develop methods to quantify redundancy and propose techniques to compare the redundancy degrees of different systems. By delving into these aspects, we seek to enhance the understanding of redundancy in complex systems and offer insights into designing highly reliable and resilient systems that can sustain continuous service provision under diverse conditions.

The main contributions of this paper are  as follows:

(1) We present a novel method to quantitatively measure independent redundancy in complex systems, taking into account the interrelations between different factors affecting redundancy. This method enables a precise evaluation of a system's ability to maintain functionality even when some of its components fail.

(2) We propose a method to calculate the Independent Redundancy Degree (IRD) when combining two systems, taking into account the independence of redundancy paths. This method allows us to determine the overall redundancy of a combined system. Meanwhile this methods can also be applied to access the IRD when remove some redundant components from a system. By utilizing this approach, we gain valuable insights into the system's resilience during system combination or separation

(3) We introduce a novel method to extend the scope of singleness factors in complex systems. These singleness factors represent factors that form a single processing path and necessitate redundant paths for improved reliability. By expanding these factors in both time and space dimensions, we can identify more critical aspects of the system that require redundancy. Addressing these extended singleness factors enables us to enhance the system's fault-tolerance and reliability by providing appropriate redundant paths.

The remaining sections of this paper are organized as follows. In Section \ref{sec_related_work}, we provide an overview of related work in the field of redundancy and fault tolerance. Section \ref{sec_IR_def} introduces the concept of Independent Redundancy Degree (IRD) and presents our method for calculating IRD when combining two systems. In Section \ref{sec_imp_facotrs}, we explore the impact factors of system redundancy. Section \ref{sec_IRD_measurement} discusses how to measure the independent redundancy degree of a system. In Section \ref{sec_verification}, we present the verification results and conduct corresponding analyses. Finally, Section \ref{sec_conclusion} concludes this paper with a summary of our contributions and potential future research directions.

\section{Related Work} \label{sec_related_work}
The exploration of system reliability and redundancy has captivated the attention of researchers and engineers across diverse domains, encompassing reliability engineering, fault tolerance, and system design. Throughout time, numerous methodologies have emerged to appraise and bolster the reliability of intricate systems. In this segment, we delve into pivotal contributions within the realm of redundancy analysis, surveying the strides taken to appraise system reliability advancements, encompassing aspects like hardware, software, and different layers of redundancy.

\subsection{Redundancy at Hardware Levels: P2P Technology and its Scope}

Redundancy strategies implemented at the hardware level have witnessed advancements, notably with the utilization of Peer-to-Peer (P2P) technology \cite{nobre2018survey}. Within the realm of hardware redundancy, P2P systems have emerged as a prominent approach. In these systems, computational tasks are distributed across peer nodes, strategically designed to reduce reliance on centralized nodes \cite{liu2021comparison}. This decentralized distribution of tasks among peers aims to mitigate the potential vulnerabilities associated with a single point of failure. Each individual peer node is equipped to function as both a server and a client, ensuring that even if a particular node encounters failure, other functional peers can seamlessly take on similar tasks. The P2P paradigm extends its utility to various domains, such as load distribution among peers \cite{liu2020distributed} and the facilitation of resource sharing among participants \cite{zia2022modeling}, encompassing activities like sharing multimedia files \cite{logeshwaran2021aicsa}.

However, it's important to note that the existing P2P approach primarily manifests as a hardware-level paradigm. While peer nodes collectively execute tasks in a distributed manner, the underlying software infrastructure often exhibits centralized features. For instance, certain P2P systems incorporate centralized software components to coordinate the functions of other software instances or to manage task scheduling, resembling traits of centralized control. Moreover, certain software implementations in P2P systems involve identical or similar software copies, such as those seen in protocols like BitTorrent \cite{erman2022bittorrent}. These characteristics collectively contribute to the persistence of centralized attributes within the software layer, even within hardware-level redundancy strategies.

\subsection{Redundancy at the Software Level: Enhancing Fault Tolerance}

In the domain of software-level redundancy, conventional strategies have primarily focused on integrating duplicate components into critical sections of a system. The central objective of these methods is to mitigate the potential impact of failures by ensuring the availability of backup components. Notable techniques include N-version programming (NVP) \cite{rodrigues2023enhancing} and Triple Modular Redundancy (TMR) \cite{barbirotta2022design}, both widely adopted to enhance fault tolerance in safety-critical systems. These approaches typically entail creating multiple redundant instances of the same software component, with their collective outputs influencing decision-making. It's worth noting that approaches like NVP and TMR often incorporate a central entity responsible for task scheduling or final computation adjustment, and the number of redundant software instances remains fixed (with N representing a fixed value, such as 2 or 3).

In more recent advancements, Su et al. \cite{su2022fully} introduced an innovative perspective by proposing independent software that operates without requiring central coordination. Furthermore, the number of redundant software instances is not predetermined, allowing new instances to join or depart, resembling a P2P software model. This independent software paradigm diverges from traditional methods, relinquishing centralized control and thereby enhancing the system's resilience to failures. These approaches have shown significant potential in bolstering system reliability. However, they may have limitations in comprehensively assessing various dimensions of redundancy, encompassing critical components like software and communication layers, which play pivotal roles in modern intricate systems. While significant progress has been made in enhancing reliability, there often remains a need for a more holistic evaluation that considers diverse facets of redundancy to ensure a robust system architecture.

\subsection{Diverse Redundancy Strategies Across Different Dimensions}

In the context of redundancy, various dimensions of a system's architecture come into play. Complex systems consist of multiple components, each susceptible to unique types of failures. Consequently, diverse redundancy technologies have emerged to address specific aspects of reliability enhancement.

For instance, to fortify power supply redundancy, the use of dual power sources has been suggested as a solution \cite{kong2019comprehensive}. This strategy ensures uninterrupted operation even if one power source fails. Similarly, to mitigate the risks posed by localized disasters, distributing machines across different geographical locations has been proposed \cite{rm2020load}. By dispersing components, the system's resilience to regional failures is amplified. To combat potential service disruptions caused by a single cloud server supplier failure, experts recommend employing multiple cloud providers \cite{tang2021reliability}. This strategy minimizes the impact of supplier-related outages.

Furthermore, the importance of communication redundancy has been highlighted. Proposals advocate for the incorporation of redundant communication networks to ensure seamless data transmission. Despite the progress in various redundancy dimensions, there remains a gap in understanding the interconnectedness of these strategies and their collective impact on overall system reliability. The relationship between these factors lacks comprehensive analysis, and methods to identify additional such factors from existing knowledge are currently absent. As systems span multiple aspects, a deeper exploration is warranted to uncover their interdependencies and to develop effective redundancy strategies that holistically address diverse dimensions of potential failure.

\subsection{Comparison}
These prior works have introduced a range of redundant methods to enhance system reliability, forming the foundation for the analysis in this paper. However, effective redundancy strategies must encompass a systematic approach to prevent potential single processing paths. These studies often overlook the exploration of interrelationships between different redundancy methods. Furthermore, assessing and comparing the quality of redundancy among various systems holds significance for guiding system design, a crucial aspect not extensively addressed in existing research.

Contrarily, our approach distinguishes itself by elucidating the interplay between different redundancy aspects and striving to identify additional factors that impact redundancy. Moreover, we introduce the concept of independent redundancy degrees to comprehensively assess system reliability. By dissecting the reliability assessment into distinct dimensions, our method facilitates a more detailed analysis of the system's performance across a spectrum of failure scenarios.

\section{Independent Redundancy in Systems} \label{sec_IR_def}
In this section, we explore the concept of independent redundancy in systems, a critical factor in ensuring system robustness and fault-tolerance. Here, a system refers to a combination of hardware, software, and other essential components responsible for processing user tasks. The term "components" is used to emphasize their role as integral parts of the entire system.

Despite the efforts to reduce errors, nearly all components in a system may be susceptible to failures, leading to potential system outages. To mitigate such risks, redundancy can be employed by incorporating multiple backup components. By doing so, even if one or some processing paths fail, the system can continue to function. Redundancy can be achieved through diverse hardware configurations  or software arrangements , creating multiple redundant paths to ensure operational reliability. For instance, in a P2P network, hardware layer redundancy can be achieved by employing different nodes, while software layers might lack redundancy.

\subsection{Definition of Independent Redundancy}
\textbf{Redundancy} refers to a system providing multiple paths to process a task. We can formally describe a redundant system using \eqref{redundancy_path_definition}, where $n$ (greater than 1) represents the total number of processing paths, and each path is capable of completing the task. These paths collectively form a set denoted as $PATH$.

\begin{equation}\label{redundancy_path_definition}
    PATH = \{ path_1, ..., path_n \}
\end{equation}
, where $path_i$ represents different processing paths, and $n$ is the total number of processing paths.

\textbf{Independent redundancy} entails the existence of various processing paths, each uniquely processing the task, rather than being copies or similar versions. As these paths are independent, they encounter errors autonomously. We can formally express this through the addition of the independent condition \eqref{independent_redundancy_path_definition} to the definition of redundancy \eqref{redundancy_path_definition}.

Redundancy is utilized to mitigate the impact of random errors that may occur in individual components or methods, as such errors cannot always be predicted. However, if redundant components are similar or identical copies of each other, they may encounter the same error. Therefore, the concept of independent redundancy focuses on avoiding similar errors among identical copies. By ensuring that redundant components are independent, the system becomes more resilient and less susceptible to widespread failures.

\begin{equation}\label{independent_redundancy_path_definition}
    \begin{array}{l}
        \exists path_i, path_j \in PATH \\
        p_{error}(path_i) \cap p_{error}(path_j) = 0
    \end{array}
\end{equation}
, where $p_{error}(path_i)$ represents the probability of $path_i$ encountering an error.
\\

The condition asserts that for any two independent processing paths $path_i$ and $path_j$ in $PATH$, their error occurrences do not overlap, ensuring autonomous error behavior.

Determining the independence of two processing paths can indeed be a challenging task. To simplify the assessment of independent redundancy, we shift the focus from independent processing paths to examining the independence of individual components. Although this method is not a necessary and sufficient condition for independent processing path(referring to section \ref{sec_challe_der_ind_path}), it is easier to implement and provides valuable insights (referring to section \ref{sec_Explanation_shift_focus_to_ind_com}). This approach introduce the concept of \textbf{independent processing components}, where each component is produced independently, meaning it is designed, implemented, and verified separately.

In the case of redundant hardware, for instance, achieving hardware independence involves utilizing CPUs of different architectures for different nodes and employing software from different manufacturers. This approach ensures that even if one hardware component fails, the redundancy provided by independent hardware elements allows the system to continue functioning. By diversifying the hardware components, the system becomes more resilient and less susceptible to widespread outages.

Consequently, we can assess the independence of two components through the correlation relationship between them. When two modules ($module_i$, $module_j$) are independent, their correlation is 0, as expressed in \eqref{eq_def_corr_moduels}.

\begin{equation}  \label{eq_def_corr_moduels}
    corr(module_i, module_j) = 0
\end{equation}

This correlation measure provides valuable information about the redundancy in the system, helping us understand the level of independence between different components and processing paths. While it may not be the sole determinant of independence, it aids in quantifying the degree of independent redundancy and its impact on system performance.

\subsubsection{Proof for the Challenging Nature of Determining Independence of Processing Paths} \label{sec_challe_der_ind_path}
(1). Different Kinds of Tasks: Consider a system with a diverse set of tasks, each with its unique requirements and processing demands. For every task, there can be numerous possible processing paths to achieve the desired outcome. Proving that two processing paths are entirely different across all possible tasks becomes infeasible due to the vast number of potential scenarios.

(2). NP-Hard Problem: The problem of determining whether two processing paths are independent for all possible tasks is classified as an NP-hard problem. NP-hard problems are known for their computational complexity, and as the number of tasks and processing paths increases, the difficulty of proving independence grows exponentially.

\subsubsection{Explanation for Shifting Focus to Independent Components} \label{sec_Explanation_shift_focus_to_ind_com}
(1). Parameters and Characteristics: Instead of attempting to prove independence for entire processing paths, we can focus on examining the parameters and characteristics of individual components involved in the system. Components, such as software binaries or design graphs of hardware, have distinct attributes that can be comparatively analyzed.

(2). Easier Implementation: Assessing the independence of individual components is relatively more manageable than trying to prove the independence of entire processing paths. This approach allows for a more practical implementation and reduces the computational complexity associated with the task.

(3). Valuable Insights: Although proving the independence of components does not guarantee independence of processing paths in all cases, it provides valuable insights into the system's redundancy and fault-tolerance capabilities. Ensuring independence at the component level enhances the system's resilience and contributes to its overall robustness.

\subsection{Redundancy and Singleness}
Redundancy involves the use of multiple components or methods to accomplish a task, ensuring a backup mechanism is available in case of failure. In contrast, singleness refers to situations where a task is achieved using only a single component or method, leaving no alternative paths in case of failure. Various manifestations of singleness can be observed:

(1). Singleness Across Different Layers: Singleness can exist at both the hardware and software layers. For hardware, having a single computer or server represents a hardware singleness scenario. Similarly, using the same or similar software across the system creates a singleness situation at the software level.

(2). Singleness within Parts of the Path: In certain cases, even when multiple hardware components are utilized, the scheduling of these components might be centralized, leading to a singleness scenario within parts of the processing path.

(3). Singleness throughout the Entire Path: Running a C program on a single computer exemplifies singleness across both hardware and software layers. Any failure, such as a hardware shutdown or software malfunction (e.g., a null pointer), will result in a complete system outage.

All contributing factors to singleness scenarios are termed \textbf{singleness factors}.

\begin{figure}
    \centering
    \includegraphics[width=3.5in]{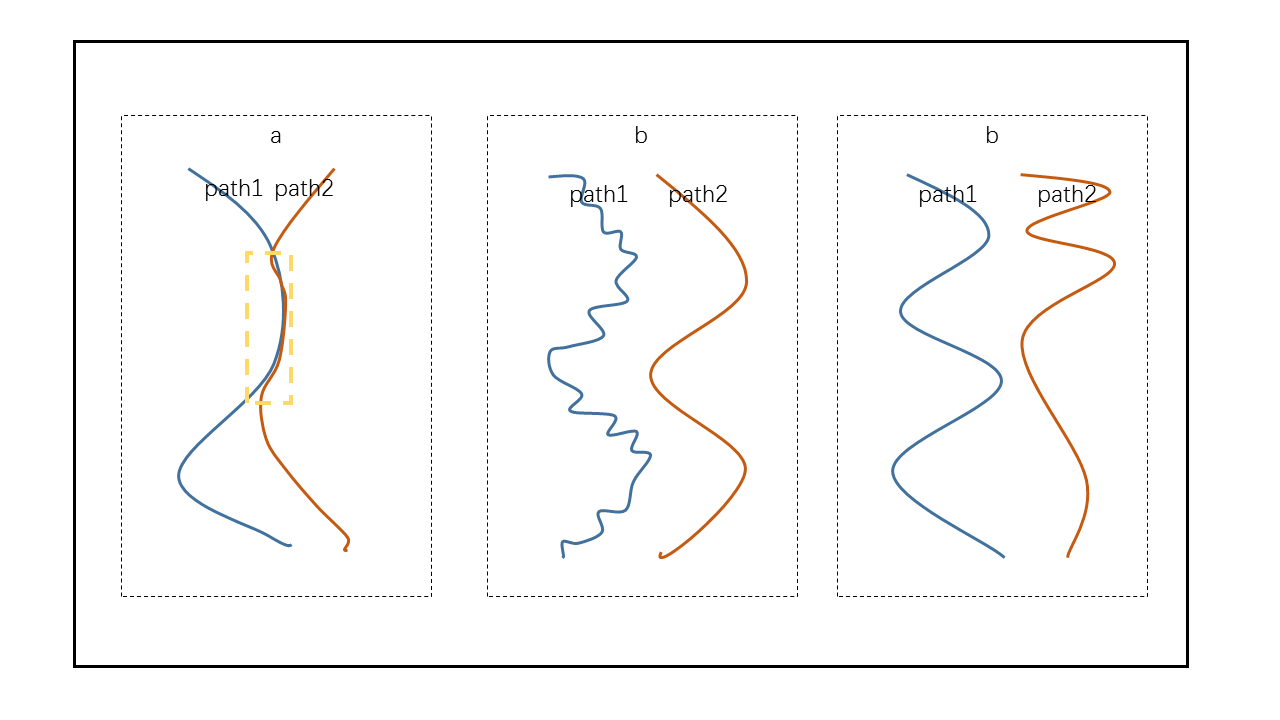}
    \caption{Different coherence of processing paths}
    \label{independentPath_demon}
\end{figure}

Figure \ref{independentPath_demon} illustrates different singleness situations. In part "a," two paths exhibit coherence as they share a segment of the processing path (in the yellow box). In part "b," while no segment is identical, they follow a similar design trend, resulting in some coherence. Part "c" displays even less coherence, with no shared processing segments or significant design similarities.

Singleness factors have the following features:

(1) Singleness factors can persist even when a system is composed of different nodes (i.e., different hardware). For instance, consider a blockchain \cite{rahardja2021immutability} comprised of different nodes. Despite having multiple nodes, the system's smart contract (software on the blockchain) may operate in a singleness format, with only one path for processing tasks. As a consequence, any code bug occurring in the smart contract would affect all nodes simultaneously, rendering the application non-functional. In such cases, there are no redundant paths to switch to, leading to a complete failure.

(2) Certain factors may exhibit redundancy at one scope while presenting singleness at another scope, such as a larger or smaller scope. The scope can refer to the space or time frame. For example, an application may utilize 100 machines to achieve hardware redundancy within the same laboratory. Despite the redundancy of multiple machines, a major natural disaster, such as a big earthquake, could damage all machines simultaneously, causing the entire application to cease functioning. In this scenario, the hardware exhibits redundancy within the laboratory scope but becomes a singleness factor when considering the broader scope of the natural disaster's impact.

\section{Impact Factors on Independence Redundancy} \label{sec_imp_facotrs}
Our aim is to achieve as high redundancy as possible. In this section, as try to analyze the factors that affect the redundancy, and how to achieve high redundancy as possible.

\subsection{Not Possible for a Full Redundancy}
A fully redundant system is one that can consistently provide at least one processing path, regardless of the condition or failures that may occur in its peers (e.g., components going out of work unintentionally or being deliberately shut down). However, achieving a truly fully redundant system is practically unattainable due to various singleness factors, some of which are unknown or unpredictable. For instance, a hacker might discover a novel method to compromise the entire system, or an unexpected earthquake could lead to the destruction of critical hardware components. These unforeseen factors introduce vulnerabilities that can undermine the attainment of complete redundancy in a system. As a result, while striving for redundancy is essential for system robustness, achieving absolute full redundancy remains a challenging goal.

\subsubsection{proof that achieving full redundancy in a system is an NP-hard problem}
To prove that the problem is NP-hard, we need to demonstrate a polynomial-time reduction from another known NP-hard problem to the full redundancy problem. The proof is to use reduction from Subset Sum Problem method.

(1). Subset Sum Problem: The Subset Sum problem is a well-known NP-hard problem. Given a set of positive integers and a target sum, the task is to determine if there exists a subset of the integers whose sum equals the target sum.

(2). Formulation of Full Redundancy Problem: We can formulate the full redundancy problem as follows: Given a system configuration consisting of components and their interconnections, we want to determine if there exists at least one processing path that remains operational under any possible failure scenario, including peer failures and shutdowns.

(3). Construction of the Reduction: To prove that the full redundancy problem is NP-hard, we will construct a polynomial-time reduction from an instance of the Subset Sum problem to an instance of the full redundancy problem.

(4). Reduction Construction: Given an instance of the Subset Sum problem with a set of positive integers and a target sum, we create a corresponding system configuration for the full redundancy problem. We create a set of components, each representing one integer from the Subset Sum problem. We also create a target processing path that corresponds to the target sum.

(5). Completeness and Correctness: The reduction ensures that the original Subset Sum problem has a solution if and only if the constructed system configuration in the full redundancy problem can provide at least one processing path in all conditions. If there exists a subset of integers whose sum equals the target sum in the Subset Sum problem, it implies that the corresponding components in the full redundancy problem can form a processing path that remains operational under any failure scenario. Conversely, if there is no such subset sum in the Subset Sum problem, it means that there is no way to create a processing path in the full redundancy problem that can provide full redundancy in all conditions.

(6). Polynomial Time Complexity: The reduction from Subset Sum to full redundancy can be performed in polynomial time. Constructing the system configuration and checking if there exists a valid processing path can be done in polynomial time with respect to the size of the input.

(7). Conclusion: Since the Subset Sum problem is NP-hard and we have shown a polynomial-time reduction from Subset Sum to the full redundancy problem, the full redundancy problem is also NP-hard.

In the above, we have provided a formal proof that achieving full redundancy in a system is an NP-hard problem using the method of reduction from the Subset Sum problem. This establishes the computational complexity of the full redundancy problem and highlights its infeasibility for finding efficient solutions for all possible system configurations and failure scenarios.
\\

Although a full independent redundency cannot achieved, we can achieve a currently available full redundancy system.
A \textbf{ currently available full redundancy system} is a system with redundency for all known singleness factors. Our goal is then to find as many singleness facotrs as possible. Although, it is difficult to find unknown factors which may be singles in a system, we can find more singleness factors from already known singleness facotrs and take according redundant methods. The methods is to use the scope related analyses method to extend more singleness facotrs from already known facotrs.

\subsection{Scope-Related Analysis of Singleness Factors}
The scope analysis involves examining factors in both the space and time dimensions.

\subsubsection{Factors Related to Different Space Scopes}
Suppose we identify that the singleness factor is caused by using a single computer. We can introduce redundancy by using several computers, but these computers might be located in a small area. We can then extend the scope by using several computers in a specified area. Each computer with its location can be regarded as an area, and these areas may be located in a small region that could be affected by a natural disaster, such as an earthquake. We can further extend the scope to different cities, and even consider different countries or planets. This illustrates that the scope can vary widely and must be carefully analyzed to ensure redundancy is effective.

The space-related singleness factors encompass the concept of different scope, which can be adjusted to identify more singleness factors. When one singleness factor is identified, its scope can be expanded or reduced to explore additional singleness scenarios. We define the concepts of \textbf{scale big} and \textbf{scale small} related to scope.

If we identify that the singleness factor is caused by using only one computer, one approach is to introduce several computers for redundancy. However, if these computers are located in a small space, such as in a laboratory, it may still result in a singleness scenario within this limited scope. To address this, we can distribute the machines to different locations, such as in different cities, to increase the scope and achieve greater redundancy. However, the scope cannot be extended indefinitely, as there should be a practical limit where the loss possibility becomes sufficiently small.

The concept of scale big can be expressed as shown in \eqref{con_scale_big}, where $SF$ represents the set of singleness factors. If factor $f$ is identified as a singleness factor, and there exists a broader factor $bf$ that includes $f$ within its scope, then $f$ belongs to the set $SF$ of singleness factors.

\begin{equation} \label{con_scale_big}
  \begin{split}
    & f \in SF \\
    & \exists bf \rightarrow  f \in bf \text{, } bf \in SF
  \end{split}
\end{equation}

The concept of scale small can be expressed as shown in \eqref{con_scale_small}, where $SF$ represents the set of singleness factors. If factor $f$ is identified as a singleness factor, and there exists a narrower factor $sf$ that is included within $f$'s scope, then $sf$ also belongs to the set $SF$ of singleness factors. For example, a computer can be considered a larger scope compared to its individual components, such as CPUs and memories. If all machines use the same CPU, they can encounter the same faults simultaneously, resulting in singleness issues at the CPU level.

\begin{equation} \label{con_scale_small}
  \begin{split}
    & f \in SF \\
    & \exists sf \rightarrow  sf \in f \text{, } sf \in SF
  \end{split}
\end{equation}
\\

By exploring different scopes, we can uncover various singleness factors that might be present at different spatial levels. Understanding these scope-related singleness factors is crucial for designing effective redundancy measures and fault-tolerant systems.

Two typical space-related singleness factors are:

(1) Layer-Related Singleness Factors: This is a special scope-related factor where a module can be divided into several layers, and each layer may introduce singleness. For example, an application can be divided into software layer and hardware layer. Redundancy in the hardware layer does not automatically provide redundancy at the software layers unless the copies of each hardware are independent.

(2) Dependence-Related Singleness Factors: Instead of altering the scope, some factors are dependence-related, meaning they depend on another known factor. For example, software identified as a singleness factor may depend on specific libraries, running virtual machines, or operating systems, which can also contribute to singleness. Similarly, in hardware, if a machine is identified as a singleness factor, its power supply could be considered a dependence-related factor. These factors exist in parallel and are not directly related to each other. This kind of related factor can be expressed as in \eqref{dependent_related_factors}.

\begin{equation} \label{dependent_related_factors}
  \begin{split}
    & \text{when } f \in SF \\
    & \exists df \rightarrow f \textbf{ dep } df \text{, and } df \in SF
  \end{split}
\end{equation}

\subsubsection{Factors Related to Different Time Scopes}
Time-related singleness factors pertain to issues that can cause all redundant processing paths to stop at the same time. One such example is when the licenses of software or hardware components expire simultaneously. Formally, we can describe time-related factors as shown in \eqref{time_singleness}. In this equation, $Redun$ represents the set of redundant components from $r_1$ to $r_n$. If all these factors have the same fault time ($fault^{time}$), they become coherent rather than independent.

For instance, if we identify that the currently used software may cause singleness, we can introduce several independent software to achieve redundancy. However, we must also consider whether the licenses of those software will expire simultaneously, potentially leading to simultaneous outages. In such cases, even with redundancy, all the redundant paths might be rendered unusable at the same time, jeopardizing the system's fault tolerance. This highlights the importance of considering time-related factors in redundancy design.

\begin{equation} \label{time_singleness}
  \begin{split}
    & \text{when } Redun = \{r_1, ..., r_i, ..., r_n\} \\
    & fault_{r_1}^{time} = ... = fault_{r_i}^{time} = ... = fault_{r_n}^{time}
  \end{split}
\end{equation}

\subsection{Redundancy Implementation Design Example: An Independent Redundancy Hardware-Software System}

In this section, we present an example of the design consideration of independent redundancy for a system composed of both hardware and software components.

Independence design of computing hardware: To achieve independent redundancy in the hardware, we need to consider the following basic requirements:
(1). Use different computing nodes with diverse types of CPUs and memories to avoid similarity in the hardware components.
(2). Ensure that computing nodes are not located on a single server to prevent a single point of failure.
(3). Ensure that the power supply of computing nodes does not depend on a single power supply line to avoid power-related singleness factors.
(4). Place the computing nodes in different cities to prevent potential simultaneous regional outages.
(5). Be cautious about the hardware's expiration date, especially in rental scenarios, to avoid all hardware becoming non-functional at the same time. This is an example of time-related factors.

Independence design of software: Achieving independent redundancy in software involves the following considerations:
(1). Use different software solutions capable of processing user tasks independently and sourced from different software companies to avoid using similar software binaries.
(2). Avoid using software that relies on the same libraries to prevent potential singleness factors at the software level.
(3). Avoid software that depends on the same runtime virtual machine to ensure independence in the software execution environment.
(4). Prefer using different operating systems for various software components to avoid operating system-related singleness factors.
(5). Be mindful of software license deadlines to prevent multiple software instances from expiring simultaneously. This is another example of time-related factors.

Independence design of communication: For communication, the following steps are crucial:
(1). Avoid using the same communication network to ensure redundancy. If one network faces an outage, the system can switch to an alternative network. The nwetwork failure of Guangdong Province for about several hours is an example.
(2) To avoid using the same network devices for all hardware or software components. For instance, if all hardware nodes are connected to a single router, a failure or shutdown of that router could lead to a complete outage of the entire system. To prevent this singleness factor, we should employ multiple routers or network devices and distribute the connections of hardware or software components across them.
(3). Use diverse types of networks, such as wired and wireless, to avoid singleness factors related to network type.
(4). Be cautious about the expiration dates of network bandwidth purchases to prevent all networks from expiring simultaneously. This is an example of time-related factors.

The above design considerations aim to address known singleness factors. However, it is essential to acknowledge that unknown factors (either yet to be discovered or unforeseen) may still exist, and achieving full independent redundancy may be challenging. Nevertheless, by implementing the described design, we can achieve a currently available independent redundancy system, which enhances system robustness and fault-tolerance.

\section{Measurement Metric: Independence Redundancy Degree (IRD)} \label{sec_IRD_measurement}

To quantify the degree of independence redundancy in a system, we propose the Independence Redundancy Degree (IRD) as a measurement metric. The IRD takes into account different singleness factors and their corresponding redundancy methods.

\subsection{Independence Redundancy Degree (IRD)}

The factors that cause a system to have a single processing path can vary, and each of these factors requires different methods to establish redundant processing paths. Therefore, we introduce dimensions ($d_i$) to differentiate different singleness factors. For the redundancy method, the value of each dimension ($v^{d_i}$) represents the probability that all redundant paths of this dimension experience an outage. The outage probability of each redundancy path ($p_{outage}$) can be determined from actual outage occurrences. If these redundancy paths are independent, the value of the dimension ($v^{d_i}$) can be expressed as shown in \eqref{eq_cal_value_di}. Thus, one dimension can be described as $v^{d_i} \times d_i$.

\begin{equation} \label{eq_cal_value_di}
  v^{d_i} = 1 - \prod \limits_{j=1}^{n} p_{outage\_j}^{d_i}
\end{equation}

The IRD of a system can then be represented as the combination of each dimension. Equation \eqref{eq_def_ird_k_m} illustrates the IRD of system $k$ with $m$ dimensions. The IRD represents the probability that there is at least one available redundant path to process a user task for each dimension, and we denote this as $d_i$, with its value referred to as the live probability of this dimension.

\begin{equation} \label{eq_def_ird_k_m}
  IRD_k = \sum_{i =1}^{m} {v^{d_i} \times \overline{d_i}}
\end{equation}

In case there is no available redundant path to process a user task (referred as outage) in dimension $k$, we can use Equation \eqref{eq_def_ird_k_m_outage}, with the value in it referred to as the outage probability. The relationship between the outage probability ($\overline{v}^{d_i}$) and live probability ($v^{d_i}$) for dimension $d_i$ is shown in Equation \eqref{eq_cal_relaition_v_v_upper}.

\begin{equation} \label{eq_def_ird_k_m_outage}
  IRD_k = \sum_{i =1}^{m} {\overline{v}^{d_i} \times \overline{d_i}}
\end{equation}

\begin{equation} \label{eq_cal_relaition_v_v_upper}
  \overline{v}^{d_i} + v^{d_i} = 1
\end{equation}

\subsection{Combination of Independent Systems}

The previous analysis focused on individual independent redundancy systems. However, there are situations where it becomes necessary to combine two or more independent redundancy systems. In such cases, calculating the Independence Redundancy Degree (IRD) of the combined system becomes crucial. Combining the values of one dimension from two IRDs poses a challenge since the paths in individual redundancy systems are independent, but when combined, their paths could become not independent.
For example, let's consider the two systems $IRD_1$ and $IRD_2$; if they place their computer hardware in the same city, the dimensions related to different cities become coherent, indicating dependency. In such cases, the combination function \textbf{cf()} can be employed to calculate the value of a dimension. The use of the combination function allows us to properly handle the dependencies and accurately determine the IRD of the combined system.

To accommodate the combination of independent redundancy systems and handle their potential interdependencies, we introduce an additional component in the IRD calculation. This component is the combination function, denoted as \textbf{cf()}. The \textbf{cf()} function has the following requirements:

(1) To ensure the correctness of the \textbf{cf()} function, it must take sufficient parameters. For instance, for geographical redundancy factors, it should accept geographical parameters as inputs.

(2) The value of the dimension is the first parameter passed to \textbf{cf()}. In \eqref{eq_def_cf_function}, it is denoted as $n$.

With the introduction of the \textbf{cf()} function, the format of the IRD changes from \eqref{eq_def_ird_k_m} to \eqref{eq_def_cf_function}. In other words, \eqref{eq_def_ird_k_m} can be considered the simplified format of \eqref{eq_def_cf_function}.

The parameters used in the \textbf{cf()} function can vary depending on the specific combination of redundancy factors. For instance, when combining redundancy factors related to geographical locations, we need to determine whether they are in the same place. On the other hand, when combining two hardware redundancy factors, their similarity should be taken into account.

The new expression for the IRD, considering the \textbf{cf()} function, is given by:

\begin{equation} \label{eq_def_cf_function}
  IRD = \sum_{i=1}^{m}{\textbf{cf$^i$}(v^{d_i}, \{p_1^i, ..., p_i^i, ..., p_n^i\})  \times d^i}
\end{equation}
, where $d_i$ represents the dimension, \textbf{cf$^i$} is the combination function for dimension $d_i$, and $v^{d_i}$ is the current value of this dimension.

Let's discuss the operations for independence redundancy degree, specifically the add and subtract operations.

\subsubsection{Add Operation of Independence Redundancy Degree}
When we combine two or more systems into a larger system, the corresponding independence redundancy degree should be adjusted accordingly. This operation is known as the add operation.

The add operation is performed one dimension at a time. The \textbf{cf()} function is utilized to calculate the final value of each dimension. In the equation \eqref{eq_add_cf_function}, we show the process of adding two IRDs, $IRD_1$ and $IRD_2$, to obtain the combined IRD $IRD_3$. In this process, both the values of each dimension and the parameters will be adjusted.

\begin{equation} \label{eq_add_cf_function}
  \begin{split}
  IRD_1 &= \sum_{i=1}^{m} {\textbf{cf$^i$}(n_1^i, \{p_1^i, ..., p_i^i, ..., p_r^i\})  \times d^i} \\
  IRD_2 &= \sum_{i=1}^{m} {\textbf{cf$^i$}(n_2^i, \{p{\prime}_1^i, ..., p{\prime}_i^i, ..., p{\prime}_s^i\})  \times d^i} \\
  IRD_3 &= IRD_1 + IRD_2  \\
        &= \sum_{i=1}^{m} {\textbf{cf$^i$}(n_3^i, \{p{\prime}{\prime}_1^i, ..., p{\prime}{\prime}_i^i, ..., p{\prime}{\prime}_t^i\}) \times d^i}
  \end{split}
\end{equation}

In the equation above, the value of each dimension in the combined IRD, $IRD_3$, is calculated using the combination function \textbf{cf()}, rather than being directly added. The parameters of the \textbf{cf()} function can be a collection of the parameters from $IRD_1$ and $IRD_2$, and some parameters may be merged. This is why the length of $IRD_3$ is denoted as $t$, which can be less than the sum of the lengths of $IRD_1$ and $IRD_2$ ($r + s$). This process ensures that the combined IRD accounts for the interdependencies and interactions between the redundancy systems being added.

The parameters used in the \textbf{cf()} function may vary depending on the specific combination of redundancy factors being considered. These parameters help identify and handle the potential dependencies between the redundancy paths, ensuring an accurate calculation of the combined IRD.

\subsubsection{Subtraction Operation of Independence Redundancy Degree}
Similarly, the subtraction operation of independence redundancy degree involves removing redundancy factors from the system, and it also utilizes the \textbf{cf()} function.

When we remove some redundancy factors from the system, such as for cost-saving purposes, the independent redundancy degree needs to be adjusted accordingly. This operation is known as the subtract operation.

Similar to the add operation, the subtract operation is carried out one dimension at a time. The \textbf{cf()} function is used to calculate the final value of each dimension in the adjusted IRD. While space constraints prevent us from presenting the detailed equations for the subtract operation here, it follows a similar approach to the add operation.

In the subtract operation, some parameters and values of dimensions may be modified or merged, depending on the specific redundancy factors being removed. The \textbf{cf()} function helps handle the interactions and dependencies between redundancy paths, ensuring an accurate calculation of the adjusted IRD after removing specific redundancy factors.

The combination and subtraction operations of independence redundancy degree are essential for assessing and optimizing redundancy strategies in complex systems. These operations allow us to compare different redundant system configurations, evaluate their effectiveness, and make informed decisions about redundancy measures based on the resulting IRD values.
\\

Lastly, the question is how to compare two independent redundancy degrees. The comparison is based on the probability that there is at least one available redundant path for each dimension.

\subsection{Comparing Independence Redundancy Degrees}

When we have two IRDs, it is crucial to compare which one has a higher redundancy level. For a system, if any dimension experiences an outage, the entire system may fail. Therefore, we aim to determine which IRD has the \textbf{minimum dimension}, rather than simply averaging the maximum dimension.

The process of comparing $IRD_1$ and $IRD_2$ can be described as follows:

(1) Initialize a variable $i$ to keep track of the comparison number and set it to 1 initially. Also, initialize a variable $max$ to record the comparison result and set it to empty initially.

(2) Check whether only one of $IRD_1$ and $IRD_2$ lacks the $i_{th}$ dimension. If so, set $max$ to the IRD that does not have the $i_{th}$ dimension. If both $IRD_1$ and $IRD_2$ lack the $i_{th}$ dimension, set $max$ to empty and proceed to step (5).

(3) If both $IRD_1$ and $IRD_2$ have the $i_{th}$ dimension, choose the dimension with the $i_{th}$ minimum value. Suppose $min_1$ corresponds to $IRD_1$, and $min_2$ corresponds to $IRD_2$. If there are multiple values with the $i_{th}$ minimum value, choose any one of them for comparison, as the other values will be used in subsequent iterations.

    (3a) If $min_1 < min_2$, it indicates that $IRD_2$ has a higher level of independent redundancy, and consequently, set $max$ to $IRD_2$. The comparison is completed, and proceed to step (5).
    (3b) If $min_1 > min_2$, it implies that $IRD_1$ has a higher level of independent redundancy, and thus, set $max$ to $IRD_1$. The comparison is completed, and proceed to step (5).
    (3c) If $min_1$ and $min_2$ are equal, then choose the second minimum value and perform the comparison until either (1) a different value is found, or (2) all values are equal.

(4) Increment $i$ by one and go back to step (2).

(5) The comparison process is concluded, and the variable $max$ contains the IRD with the higher level of independent redundancy.

\begin{algorithm}[h]
  \caption{Comparison of $IRD_1$ and $IRD_2$}\label{algorithm1}
  \begin{algorithmic}[1]
      \STATE Initialize $i \gets 1$ and $max \gets \text{empty}$
      \WHILE{Not all dimensions compared}
          \IF{Only one of $IRD_1$ and $IRD_2$ lacks $i_{th}$ dimension}
              \STATE Set $max$ to the IRD without the $i_{th}$ dimension
          \ELSIF{Both $IRD_1$ and $IRD_2$ lack $i_{th}$ dimension}
              \STATE Set $max$ to \text{empty}
          \ELSE
              \STATE Find $i_{th}$ minimum values $min_1$ and $min_2$ for $IRD_1$ and $IRD_2$, respectively
              \IF{$min_1 < min_2$}
                  \STATE Set $max$ to $IRD_2$
              \ELSIF{$min_1 > min_2$}
                  \STATE Set $max$ to $IRD_1$
              \ELSE
                  \STATE continue
              \ENDIF
          \ENDIF
          \STATE Increase $i$ by one
      \ENDWHILE
      \STATE \textbf{End}
  \end{algorithmic}
  \end{algorithm}

This algorithm allows us to determine which system has a higher level of independent redundancy, aiding in decision-making when designing and optimizing redundant systems.

\section{Verification}\label{sec_verification}
In this section, we try to simulate a system with hardware, software and communication redundancy to verify results of different redundency methods.

\subsection{Outage affected By different Dimensions }
\subsubsection{Environment} \label{sec_ver_en_first}
1) Hardware Simulation

In this system, there are 30 computers distributed across 3 labs in a city. The entire system fails to function properly if all the computers are out of order. The system outage is influenced by three distinct layers:

Layer 1: Each individual computer has a probability ($p_{single}$, 0.5) of experiencing an outage, which could be due to reasons like hardware faults or being intentionally shut down. If all 30 computers have been affected by an outage, the probability of which is given by $p_{single}^{30}$, then the entire system will not work.

Layer 2: Similarly, each lab has a probability ($p_{lab}$, 0.1) of not working, which could be due to reasons like lab-wide technical issues or hazards like fires. If all 3 labs are affected by non-functionality, the probability of which is given by $p_{lab}^{3}$, then the entire system will not work.

Layer 3: Beyond individual labs, the entire city also has a probability ($p_{city}$, 0.01) of being in a state of disaster, which could involve events like earthquakes. If the city is in a disaster, the probability of which is given by $p_{city}$, then the entire system will not work.

To summarize, the system's functionality relies on the successful operation of its individual computers (Layer 1), the proper functioning of each lab (Layer 2), and the absence of a city-wide disaster (Layer 3). If any of these layers fail, the whole system will not work.

2) Software Simulation

Due to cost considerations in each project, only three independent redundant software solutions are employed. The system outage is influenced by two distinct layers:

Layer 1: Each individual software has a probability ($p_{logic}$, 0.1) of experiencing an outage, which could result from reasons such as code errors or code vulnerabilities. If all three software instances are affected by an outage, with a combined probability of $p_{logic}^{3}$, then the entire system will fail to operate.

Layer 2: Additionally, each individual software has a probability ($p_{time}$, 0.01) of outage due to time-related factors, like software license expiration. If all three software instances have expired, with a combined probability of $p_{time}^{3}$, then the entire system will fail to function.

3) Communication Simulation

The computers are distributed across 3 labs, and the system outage is influenced by three distinct communication-related layers:

Layer 1: Assuming each lab has an outgoing router, each router has a probability ($p_{router}$, 0.05) of experiencing an outage, which could result from reasons like hardware faults or intentional shutdowns. If all 3 routers are affected by an outage, with a combined probability of $p_{router}^{3}$, then the entire system will not work.

Layer 2: Since the labs are located within one city, the entire system has the potential to experience an outage if the network in the whole city is affected, with a probability of $p_{city}$ (0.001). For instance, the outage of China Telegram in Guangdong Province serves as an example.

Layer 3: Additionally, each individual lab has a probability ($p_{time\_network}$, 0.01) of experiencing an outage due to time-related factors, such as software license expiration. If all 3 network instances have expired, with a combined probability of $p_{time\_network}^{3}$, then the entire system will fail to function.

These probabilities cannot always maintain at the same value, it may varies. To simulate this, we  give a variety on the probability within the range of 0.1.
\subsubsection{Results and Analysis}
We conducted four different test scenarios to explore the impact of different dimensions on system outages. The objective was to observe how the presence or absence of outages in specific layers affects the overall system's performance and reliability. By analyzing these scenarios, we gained insights into the contribution of individual dimensions to the system's resilience and availability.

Case 1 ('All Possible Outage'): In this scenario, outages occurred in all aspects, including hardware, software, and communication layers.

Case 2 ('No Hardware Outage'): This scenario simulated the absence of hardware outages. Outages occurred in the software and communication layers, but there were no hardware outages.

Case 3 ('No Software Outage'): This scenario simulated the absence of software outages. Outages occurred in the hardware and communication layers, but there were no software outages.

Case 4 ('No Communication Outage'): This scenario simulated the absence of communication outages. Outages occurred in the hardware and software layers, but there were no communication outages.

\begin{figure}
  \centering
  \includegraphics[width=3.5in]{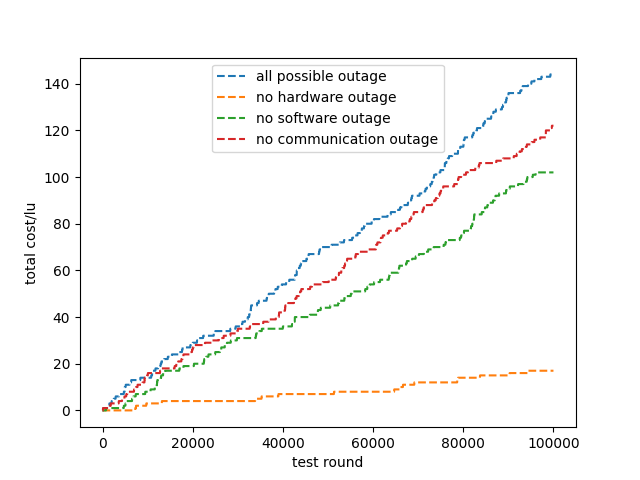}
  \caption{The outage comparison.}
  \label{different_outage}
\end{figure}

We performed 100,000 tests, i.e., running the system 100,000 times. In each test round, each component generated a random number ($p_{component}$, between 0 and 1). If $p_{component}$ was less than the outage probability as described in \eqref{sec_ver_en_first}, the corresponding component experienced an outage. The total number of outages for the entire system is shown in Figure \ref{different_outage}.

From Figure \ref{different_outage}, we observed that with an increase in the number of test rounds, the number of outages also increased. Despite some fluctuations, the overall trend appeared linear. The "All Possible Outage" case had the highest outage count. For example, there were 144 system outages, whereas there were only 17, 102, and 122 outages for the "No Hardware Outage", "No Software Outage", and "No Communication Outage" cases, respectively, in the 100,000th test round. Moreover, the sum of the outage counts for the individual cases was more than twice the outage count of the "All Possible Outage" case. This was because multiple outages in different aspects (e.g., hardware and software) occurred during a single test round, resulting in only one system outage.

The "No Hardware Outage" line consistently lay below the other three lines, indicating that the system outages were mainly caused by hardware issues. For instance, in the 60,000th test round, there were only 8 outages in the "No Hardware Outage" case, whereas there were 54, 69, and 81 outages in the "No Software Outage", "No Hardware Outage", and "No Communication Outage" cases, respectively.

If one layer is not considered, the number of simulated outages is bigger than the actual outage counts. Thus, it is crucial to consider as many outage factors as possible. Conversely, if there are more system outages observed than expected, it may indicate the existence of unknown aspects that are not redundantly addressed.

\begin{figure}
  \centering
  \includegraphics[width=3.5in]{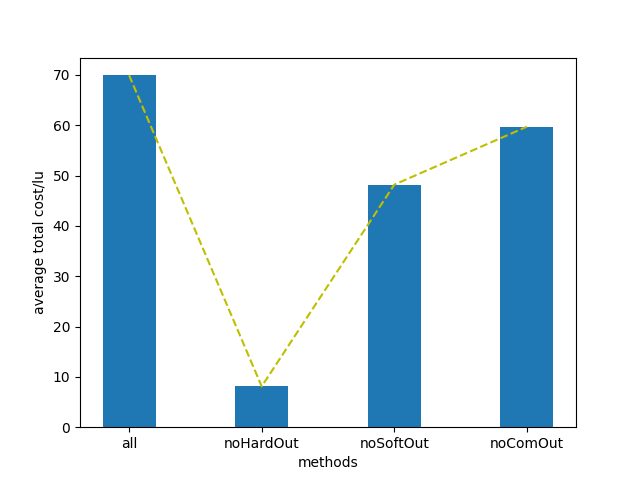}
  \caption{The average outage comparison.}
  \label{different_outage_bar}
\end{figure}

To further demonstrate the impact of different dimensions, we show the total outage counts in Figure \ref{different_outage_bar}. It is evident that the "No Hardware Outage" case has significantly fewer outages compared to the other cases. This case corresponds to the dimension with the least IRD, and applying the redundancy method to minimize this IRD can greatly improve the system's resilience to outages.

\subsection{Analyzing Dimensional Weakness}
In this section, we aim to demonstrate the concept that the first dimension to be broken is the one that minimizes the Independent Redundancy Degree (IRD). To verify this, we conducted several test scenarios using different IRD instances. These instances, denoted as $IRD_1$, $IRD_2$, $IRD_3$, up to $IRD_6$, are shown in \eqref{eq_def_four_irds}, with variations in dimension values $\overline{d_1}$, $\overline{d_2}$, and $\overline{d_3}$.

\begin{equation} \label{eq_def_four_irds}
\begin{aligned}
  IRD_1 &= 0.001  * \overline{d_1} + 0.005 * \overline{d_2} + 0.009 * \overline{d_3} \\
  IRD_2 &= 0.002  * \overline{d_1} + 0.005 * \overline{d_2} + 0.008 * \overline{d_3} \\
  IRD_3 &= 0.003  * \overline{d_1} + 0.005 * \overline{d_2} + 0.007 * \overline{d_3} \\
  IRD_4 &= 0.004  * \overline{d_1} + 0.005 * \overline{d_2} + 0.006 * \overline{d_3} \\
  IRD_5 &= 0.0049 * \overline{d_1} + 0.005 * \overline{d_2} + 0.0051* \overline{d_3} \\
  IRD_6 &= 0.005  * \overline{d_1} + 0.005 * \overline{d_2} + 0.005 * \overline{d_3}
\end{aligned}
\end{equation}
, where $\overline{d_1}$, $\overline{d_2}$, and $\overline{d_3}$ are three dimensions, and the dimensions with the overlines indicate their values represent the outage probabilities.

To simulate the system's operation, we ran it 10000 times until a dimension fails. The simulation involved generating random numbers (ranging from 0 to 1) using Python's 'random' function, representing the real outage probabilities.
(1) If the real probability is within the outage range (e.g., for an outage probability of 0.005, the real outage probability is less than 0.005), the corresponding dimension experiences an outage, and the value of 'runAccount' is logged into a separate file (the filename is related to the name of each IRD). The result data is obtained from this file.
(2) Otherwise, the system continues to run normally, and 'runAccount' increases by 1.

\begin{figure}
  \centering
  \includegraphics[width=3.5in]{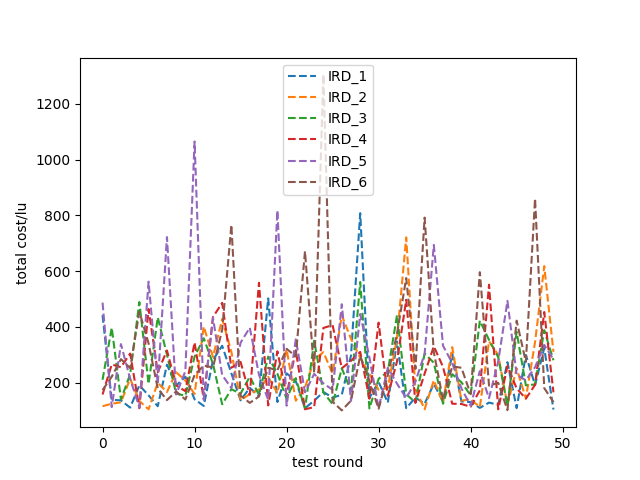}
  \caption{The different success probability of different IRD values}
  \label{ana_di_wek}
\end{figure}

The results of the first 50 test rounds are shown in Figure \ref{ana_di_wek}. From this Figure, we observe that $IRD_6$, $IRD_5$, and $IRD_4$ have the highest peaks. For instance, in the 24$_{th}$ test round, the system with $IRD_6$ method ran successfully 1306 times, and in the 10$_{th}$ test round, the system with $IRD_5$ method ran successfully 1065 times. In contrast, the remaining three IRDs did not exhibit high values of successful runs.

\begin{figure}
  \centering
  \includegraphics[width=3.5in]{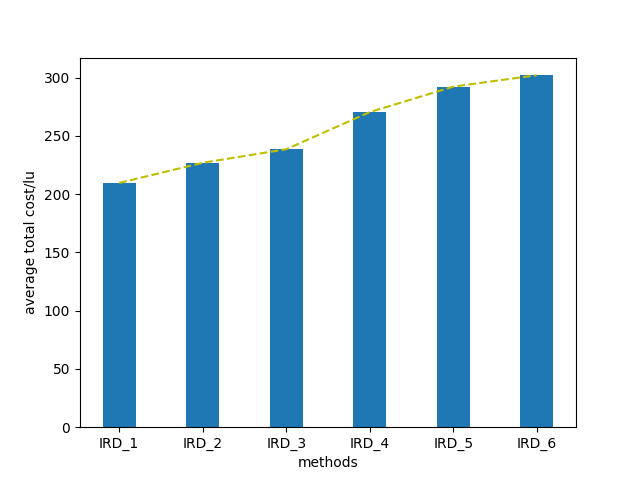}
  \caption{The different success probability of different IRD values}
  \label{ana_di_wek_bar}
\end{figure}

From Figure \ref{ana_di_wek}, we can observe that as the difference between IRD values decreases, the number of successful runs increases. To further illustrate this trend, we show the average values in Figure \ref{ana_di_wek_bar}. This figure clearly depicts the trend of successful probability. As the difference between dimensions decreases (even though the sum of their outage probabilities remains the same at 0.15), the successful probability increases. This suggests that we should strive to equalize the values of each dimension to improve system reliability.

\subsubsection{Probability to Reach Enough Low Outage Probabilities}
In this section, our objective is to explore the impact of increasing the number of independent redundancy modules on the system's outage probability in one dimension. We aim to determine how the outage probability of the entire system changes with varying numbers of independent redundancy modules.

The probabilities of normal operation ($p_{\text{success}}$) range from 0.9 to 0.1 in steps of 0.1. The number of independent redundancy modules ($\text{number}$) varies from 1 to 70. The success probability of the entire dimension ($p_{\text{success}}^{\text{dimension}}$) is the complement of the probability that all independent redundancy modules function incorrectly, as represented in \eqref{eq_cal_p_outage}.

\begin{equation} \label{eq_cal_p_outage}
  p_{\text{success}}^{\text{dimension}} = 1 - \prod_{i=1}^{\text{number}}(1 - p_{\text{success}})
\end{equation}

The equation calculates the outage probability by taking the product of the probabilities of individual modules working normally (each $(1-p_{\text{success}})$ represents the probability of one module failing) and then subtracting this value from 1 to get the overall outage probability.

\begin{figure}
  \centering
  \includegraphics[width=3.5in]{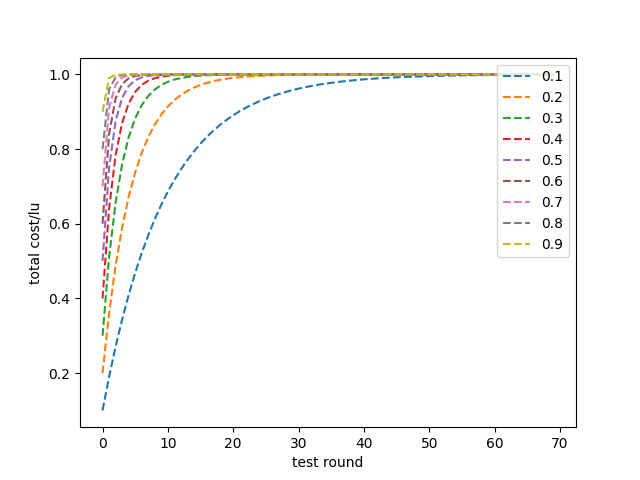}
  \caption{The impact of different number of independent redundency modules}
  \label{impactOfIncreaseOfIDModules}
\end{figure}

The results are presented in Figure \ref{impactOfIncreaseOfIDModules}. From this figure, we observe that there are distinct trends with the increase in the number of independent redundancy modules. Initially, as the number of independent redundancy modules increases, the probability of the entire system operating successfully experiences a significant improvement. However, in the later stages, further increasing the number of independent redundancy modules does not lead to substantial gains in the probability of the entire system operating successfully.

For instance, when the individual module success probability ($p_{\text{success}}$) is 0.5, the system success probability ($p_{\text{success}}^{\text{system}}$) increases to 0.992 with 6 independent redundancy modules, and further increases only slightly to 0.999 with 9 independent redundancy modules. Furthermore, even for the probability of success is only 0.1 for a single module, when there are 30 independent redundancy modules, the success probability of the system is more than 90\%.

Figure \ref{impactOfIncreaseOfIDModules} indicates that deploying an excessive number of independent redundancy modules may not be necessary. By strategically selecting a certain number of redundancy modules, it is possible to achieve a sufficiently high probability of success while saving resources on deploying additional modules that do not significantly impact the success probability. This approach helps strike a balance between the desired probability of success and the associated cost.

\section{Conclusion} \label{sec_conclusion}
This paper addresses the concept of systematic redundancy within application systems. We formally introduce the concept of Independent Redundancy Degree (IRD) as a means to quantify the redundancy levels of diverse systems, taking into account various impact factors, spanning hardware, software, and communication domains. We further delve into singleness factors, which necessitate redundancy processing paths, and propose a method to extend the scope of these factors, allowing the identification of additional singleness factors using temporal and spatial perspectives.
Moreover, our study encompasses the analysis of IRD operations and introduces a methodology for comparing two IRDs. Through verification, we present the results that illustrate the effects of different dimensions and the implications of weakness dimensions. This research contributes to enhancing our understanding of redundancy design, system resilience, and fault-tolerance by providing a comprehensive framework for evaluating and quantifying redundancy in complex systems.

\section*{Acknowledgment}
The authors thank the anonymous reviewers for their constructive comments, which help us to improve the quality of this paper. This work was supported in part by the National Natural Science Foundation of China under Grant No. 61772352; the Science and Technology Planning Project of Sichuan Province under Grant No. 2019YFG0400, 2018GZDZX0031, 2018GZDZX0004, 2017GZDZX0003, 2018JY0182, 19ZDYF1286.

\ifCLASSOPTIONcaptionsoff
  \newpage
\fi



%

\bibliographystyle{unsrt}
\bibliography{paperbib}

\begin{thebibliography}{10}

\bibitem{ma2019high}
Zheng Ma, Ming Xiao, Yue Xiao, Zhibo Pang, H~Vincent Poor, and Branka Vucetic.
\newblock High-reliability and low-latency wireless communication for internet
  of things: Challenges, fundamentals, and enabling technologies.
\newblock {\em IEEE Internet of Things Journal}, 6(5):7946--7970, 2019.

\bibitem{elbamby2019wireless}
Mohammed~S Elbamby, Cristina Perfecto, Chen-Feng Liu, Jihong Park, Sumudu
  Samarakoon, Xianfu Chen, and Mehdi Bennis.
\newblock Wireless edge computing with latency and reliability guarantees.
\newblock {\em Proceedings of the IEEE}, 107(8):1717--1737, 2019.

\bibitem{li2019cost}
Xiaole Li, Hua Wang, Shanwen Yi, and Linbo Zhai.
\newblock Cost-efficient disaster backup for multiple data centers using
  capacity-constrained multicast.
\newblock {\em Concurrency and Computation: Practice and Experience},
  31(17):e5266, 2019.

\bibitem{nobre2018survey}
J{\'e}ferson~Campos Nobre, Cristina Melchiors, Clarissa~Cassales Marquezan,
  Liane Margarida~Rockenbach Tarouco, and Lisandro~Zambenedetti Granville.
\newblock A survey on the use of p2p technology for network management.
\newblock {\em Journal of Network and Systems Management}, 26:189--221, 2018.

\bibitem{liu2021comparison}
Congcong Liu and Zhengshuo Li.
\newblock Comparison of centralized and peer-to-peer decentralized market
  designs for community markets.
\newblock {\em IEEE transactions on industry applications}, 58(1):67--77, 2021.

\bibitem{liu2020distributed}
Hong Liu, Jifeng Li, Shaoyun Ge, Xingtang He, Furong Li, and Chenghong Gu.
\newblock Distributed day-ahead peer-to-peer trading for multi-microgrid
  systems in active distribution networks.
\newblock {\em IEEE Access}, 8:66961--66976, 2020.

\bibitem{zia2022modeling}
Kashif Zia, Arshad Muhammad, and Dinesh~Kumar Saini.
\newblock Modeling and assessment of resource-sharing efficiency in social
  internet of things.
\newblock In {\em Dependable IoT for Human and Industry}, pages 83--103. River
  Publishers, 2022.

\bibitem{logeshwaran2021aicsa}
J~Logeshwaran.
\newblock Aicsa-an artificial intelligence cyber security algorithm for
  cooperative p2p file sharing in social networks.
\newblock {\em ICTACT Journal on Data Science and Machine Learning},
  3(1):251--253, 2021.

\bibitem{erman2022bittorrent}
David Erman, Dragos Ilie, and Adrian Popescu.
\newblock Bittorrent session characteristics and models: Extended version.
\newblock In {\em Traffic and Performance Engineering for Heterogeneous
  Networks}, pages 61--84. River Publishers, 2022.

\bibitem{rodrigues2023enhancing}
J{\'u}lio Rodrigues~de Mendon{\c{c}}a~Neto, Fumio Machida, and Marcus Volp.
\newblock Enhancing the reliability of perception systems using n-version
  programming and rejuvenation.
\newblock {\em Enhancing the Reliability of Perception Systems using N-version
  Programming and Rejuvenation}, 2023.

\bibitem{barbirotta2022design}
Marcello Barbirotta, Abdallah Cheikh, Antonio Mastrandrea, Francesco
  Menichelli, and Mauro Olivieri.
\newblock Design and evaluation of buffered triple modular redundancy in
  interleaved-multi-threading processors.
\newblock {\em IEEE Access}, 10:126074--126088, 2022.

\bibitem{su2022fully}
Hong Su, Bing Guo, Junyu Lu, and Xinhua Suo.
\newblock Fully decentralized model by p2p smart contract to achieve high
  availability in iot.
\newblock 2022.

\bibitem{kong2019comprehensive}
Weiwei Kong, Yugong Luo, Zhaobo Qin, Yunlong Qi, and Xiaomin Lian.
\newblock Comprehensive fault diagnosis and fault-tolerant protection of
  in-vehicle intelligent electric power supply network.
\newblock {\em IEEE Transactions on Vehicular Technology}, 68(11):10453--10464,
  2019.

\bibitem{rm2020load}
Swarna~Priya RM, Sweta Bhattacharya, Praveen Kumar~Reddy Maddikunta, Siva
  Rama~Krishnan Somayaji, Kuruva Lakshmanna, Rajesh Kaluri, Aseel Hussien, and
  Thippa~Reddy Gadekallu.
\newblock Load balancing of energy cloud using wind driven and firefly
  algorithms in internet of everything.
\newblock {\em Journal of parallel and distributed computing}, 142:16--26,
  2020.

\bibitem{tang2021reliability}
Xiaoyong Tang.
\newblock Reliability-aware cost-efficient scientific workflows scheduling
  strategy on multi-cloud systems.
\newblock {\em IEEE Transactions on Cloud Computing}, 10(4):2909--2919, 2021.

\bibitem{rahardja2021immutability}
Untung Rahardja, Achmad~Nizar Hidayanto, Ninda Lutfiani, Dyah~Ayu Febiani, and
  Qurotul Aini.
\newblock Immutability of distributed hash model on blockchain node storage.
\newblock {\em Sci. J. Informatics}, 8(1):137--143, 2021.

\end{thebibliography}

%




\begin{IEEEbiographynophoto}{Hong Su}
  Su Hong received the BS, MS and PHD degrees, in 2003, 2006 and 2022, respectively, from Sichuan University, Chengdu, China. He is currently an associate professor in Chengdu University of Information Technology, Chengdu, China. His research interests include blockchain and Internet of Value.
\end{IEEEbiographynophoto}

\end{document}